# Universal Behavior of Opponent Statistics and Applications to the MLB

Francis Liu
fliu5@mit.edu

**Abstract**

In most popular sports leagues, like the MLB, NBA, and NFL, none of the commonly used statistics take into account the strengths of the opponents a player faces. One of the main reasons for this is the conventional belief that a player's luck tends to even out over the course of a season. The other main reason is the difficulties of finding a sensible algorithm to both quantify the strengths of the opponents and incorporate such quantifications into a renormalization of a player's statistics.

In this paper, we first argue that certain statistics, such as Earned Run Average (ERA) or Fielding Independent Pitching (FIP) can be significantly skewed by opponents' strengths in the MLB. We then present an algorithm to renormalize such statistics, using FIP as the main example. This is achieved by observing that certain opponent statistics for all 30 teams in the MLB (e.g. the collection of each game's opponent FIP value over the course of a season) follow a universal distribution, up to scaling and shift.

This enables us to establish a data set for a hypothetical average team and to develop a pitching statistic based on FIP which accounts for the strength of a pitcher's schedule through methods based on equipercentile equating. It is called aFIP, which measures what a pitcher's FIP would have been if he had faced a league-average offensive team every time he pitched.

We find that there is a significant difference between aFIP and FIP for some pitchers during the 2019 season and other seasons as well, adding a new tool for player evaluation. This could make millions of dollars of difference in player contracts and in profits for teams as they enhance the accuracy with which they make player acquisitions. The universal distribution we observed also has many possible future applications throughout the sports world.

# 1 INTRODUCTION

Modern sports are built on the evaluation of players. Maximizing the amount of talent a team can acquire with their available financial resources is key to their success. There is a great amount of money at play. For example, player contracts in the MLB range from $575,500 to over $37M, while team values range from $990M to $5.3B.

Every front office member, agent, fan, and analyst wants to be able to accurately determine which player or team is superior, and by how much. A large number of statistics have been invented for this purpose and are constantly being improved. These statistics measure a player's abilities from different aspects as well as their values to a team. For example, in baseball, pitchers are evaluated using statistics like ERA (Earned Run Average) and FIP (Fielding Independent Pitching). ERA is the average number of earned runs a pitcher gives up per nine innings, which is the most obvious and fundamental to evaluate a pitcher's ability, as his job is to keep as few runs from scoring as possible. However, ERA can be a biased statistic because there are many factors contributing to it that are out of the pitcher's control, such as his fielders' defensive abilities and just random luck in where the ball is hit in relation to where the fielders are positioned.

FIP takes away these factors. FIP calculates what a pitcher's ERA would be if they had an average Batting Average on Balls in Play (BABIP). FIP is calculated only using walks, hit by pitch, strikeouts, and home runs -- factors that are not impacted by what happened after the ball was put in play and in the hands of luck and the fielders' abilities. [1] While ERA and FIP measure a player's performances, WAR (Wins Above Replacement) was invented to compare players' overall contributions to their teams by evaluating how many more wins a player brought his team than an average replacement player would. Using WAR, players of different positions can be compared. [2][3]

Similar things can be said about other sports such as professional basketball and football: there are statistics for positions or skills as well as WAR-type stats to measure overall contribution to the team.

Individual statistics have come a long way and have become essential tools for player comparison. There have been significant developments for team evaluation as well. Teams were originally just evaluated on their record, and sometimes their run differential as well. But new ways of evaluation have been developed and used in recent years. One of those methods is Pythagorean Win-Loss, a method of using the total amount of runs scored and runs allowed by a team to predict what their record would be were it not for good luck in close games(a team that loses a game 5-4 and wins a game 13-2 is not the same as a team that wins a game 2-1 and loses a game 8-0). BaseRuns is a similar metric(and was also used), except it estimates the number of runs a team would be expected to score or allow given their offensive performance without context. [4][5]

In most popular sports leagues, like the MLB, NBA, and NFL, none of these commonly used statistics take into account the strengths of the opponents a player or a team faces. Intuitively, the strengths of teams a player faces should play a role in a player's evaluations. In many other sports and competitions, the opponents' strengths play a key role in how much credit a team gets for a win or loss. One example is chess, in which the amount a player's rating changes is dependent on their opponent's rating. It should be that way -- a player should get more credit for beating a strong opponent than a weak opponent, and should not be penalized as much if they lose to a high-level player. It would be misleading to treat every game outcome equally. [6]

Similar ideas are used in the SAT and many other standardized tests. A student's final score is adjusted depending on the difficulty of the test. Missing two questions may result in a deduction of 20 points on one test, and 30 on another. This is only fair -- missing only two questions on a hard test is more impressive than missing two questions on an easier test, and so the final score should reflect that. In order to properly evaluate which tests are easier and harder, by how much, and to decide the extent to which a certain test's difficulty should impact the scoring curve, the analysts must examine how all students did on the test. If one thinks of an individual student as similar to a player or team in sports, and the test as the opponent, then the SAT scoring system is a method of evaluating players or teams using opponent statistics -- just like how it is done in chess. [7]

A crucial difference between chess and the MLB, NBA, or NFL is that the former is an individual sport. In chess, there are many opponent players, and if a player is winning, they will keep playing stronger and stronger opponents. But in the MLB, NBA, and NFL there are only 30-32 teams. In an MLB season, each team plays 162 games, and it is generally expected that everyone will face roughly the same amount of strong and weak teams. Thus, it is often argued that there is no need to account for the opponents' relative strengths because they even out over the course of the season.

This is, however, not always the case. In the MLB, the schedule is such that almost half of a team's games are against division opponents. Organizations in divisions stacked with contending teams face stronger opponents more often. This is particularly pronounced for pitchers, who have at most 30-33 appearances in a season. Among these outings, the strengths of a pitcher's opponents can be significantly skewed. For example, in the 2021 season, a pitcher for the Orioles will have to face the potent offenses of the Red Sox, Rays, Yankees, and Blue Jays on many occasions. If he gets unlucky on which days his starts happen to land, he could get over half of his starts against tough lineups. If he gets even more unlucky, he could draw starts against tough teams outside his division too. If it happens to be his turn to pitch when the Dodgers come to town or when his team flies out to Houston to play the Astros, he could be facing a far tougher challenge during the season than many other pitchers. In contrast, a pitcher for the White Sox gets to match up against the rebuilding Twins, Tigers, and Royals many times.

Take Jose Berrios and Charlie Morton in 2019 as an example. Berrios faced seven top 10 run-scoring teams and three top 5 teams while facing seventeen bottom 10 scoring teams and seven bottom 5 teams. Morton faced four bottom 5 teams and twelve bottom 10 teams while facing sixteen top 10 teams and thirteen top 5 teams. Berrios clearly had a much easier schedule than the average pitcher, while Morton had a harder one. As a result, directly comparing the FIP of two pitchers could be misleading.

*Table 1: Schedule Comparison Between Jose Berrios and Charlie Morton*

| | Berrios | Morton |
|---|---|---|
| T5 | 3 | 13 |
| T10 | 7 | 16 |
| B10 | 17 | 12 |
| B5 | 7 | 4 |
| Total Games Pitched | 32 | 33 |

Similar situations occur in the NFL as well, considering that teams play just 16 games -- six of them being against division opponents and four of them being against a different designated division. For example, in 2020 New York Jets quarterback Sam Darnold played 12 games, among which only two were bottom 10 scoring defenses and one was bottom 5. Meanwhile, Aaron Rodgers from the Green Bay Packers played 16 games and faced seven bottom 10 scoring defenses and five in the bottom 5.

In these situations, current statistics, which do not reflect unevenness in a player's opponents, could be misleading. Much is at stake. Small differences, such as a 10% change in a pitcher's ERA or FIP, could significantly change people's perceptions of his ability resulting in differences in playing time, trades, contracts, and awards. In Darnold's case, the Jets used the 2020 season to help decide whether or not he could be their franchise quarterback. The answer was no, and he was traded in the offseason.

Even if one wants to take into account the strengths of opponents in statistics for, say, the MLB, there is no obvious way to do it. As mentioned before, with individual events the strength of an opponent is much easier to quantify. In chess, the opponent's ability is simply quantified as their rating. For the SAT, the difficulty of the test is reflected by all students' raw scores (mean, median, standard deviation, etc.).

In team sports, team statistics are often used to rank the teams, but cannot be used to put quantitative values on the strength of a team. For example, the 2018 Red Sox can be considered the best offensive team in baseball because they scored the most runs. If we give their offense a score of 100, how would we determine a score for the Yankees, who scored the second-most runs? The immediate response may be just to use runs scored. Since the Red Sox scored 876 runs and the Yankees 851, we may assign the Yankees a score of 100*(851/876) = 97.15. But this is problematic.

For the moment, let's assume that a score can be assigned to the Yankees. How can one use each team's scores to evaluate players with the opponent's strengths taken into account? Also, as mentioned before, simply assigning scores or just going off runs scored is insufficient. Imagine a scenario where the Red Sox had a very boom or bust offense. 30% of the time, they would score just 2 or fewer runs off the starting pitcher. However, in the other 70% of the time, they often put up such high run totals that they were propelled into the #1 spot. Now suppose the Yankees were very consistent. Over 90% of the time, they scored more than two runs off the starting pitcher, but they rarely put up a very high run total. If a pitcher gives up two runs to the Red Sox, it can be considered a pretty solid start. If he were able to do that against the Yankees, it intuitively appears that he pitched even better that day. Clearly, such nuances cannot be captured by assigning these teams simple offensive scores. More sophisticated tools are needed.

In this paper, we show that for certain statistics, such as ERA or FIP, such difficulties can be circumvented and statistical meaningful "renormalizations'' can be applied to their "raw" statistics to reflect the strengths of their opponents. If we use SAT as an analogue, the FIP of a pitcher against a team may be considered as his "raw" score. We just need to create a procedure to convert it to a "final" score with the opponents' strength taken into account.

The key lies in a surprising observation of universal behavior of the curve of a team's opponent statistics. More explicitly, if we plot the histogram of the opponent pitchers' FIPs from each game for all 30 MLB teams, we obtain almost exactly the same curve for each team up to a horizontal shift or overall scaling. In other words, the shapes of the curves are the same for all teams. The differences among the team lie in the horizontal shifts or overall scalings which encode strengths of the teams. We can then combine all the data points to create a hypothetical league average team, which can then be used as the "standard ruler'' with which FIPs of different pitchers against different teams can be compared and renormalized.

This enables us to quantify the extent to which strength of schedule impacted a pitcher's performance by introducing the statistic aFIP -- predicting what a pitcher's FIP would have been had he faced a league-average team every time out. For example, if a pitcher had a 4.00 FIP against a team, putting it through our algorithm yields a renormalized number, say 4.12. This way, we have determined that if he had faced an average team that day, he would likely have had a 4.12 FIP instead. His aFIP for that start is therefore 4.12. One can do this for each of a pitcher's starts and calculate his season FIP.

We will use FIP as the main example to illustrate the idea. The same idea can be used for improving other statistics as well. We have also observed similar universal behavior in opponent statistics for other leagues such as the NBA and NFL, and the same idea can be used for improving other statistics as well.

The universality we observed also has other applications. When a team's curve deviates significantly from the universal curve, there may be systematic reasons behind it. A possibility is artificial tampering.

For example, the 2019 Houston Astros were being investigated for cheating in home games. We examine a possible indication of artificial tampering, by seeing whether their opponent FIP data sets for home and away games still had the same distribution.

# 2 EXTRACTING DATA

FIP calculates what a pitcher's ERA would be if they had an average Batting Average on Balls in Play (BABIP). This is because what happens after the batted ball is put in play is not a reflection of a pitcher's ability. Pitchers can suffer from a difference in results because of the ability of the fielders or just random luck causing balls to get hit directly at fielders more or less often. According to Fangraphs' definition: "the amount of balls that fall in for hits against pitchers do not correlate well across seasons" [8]. In other words, pitchers have little control over balls in play, and short-term fluctuations in BABIP are not attributable to the pitcher.

As mentioned before, FIP is calculated only using the factors that can be controlled by the pitcher -- strikeouts, walks, hit by pitch, and home runs. FIP is still flawed because there are a handful of pitchers who are better at inducing weak contact on balls in play than others, but it is a far less random metric than ERA. FIP is calculated like so: [8][9]

$$(homeruns \times 13 + walks/hit\ by\ pitch \times 3 - strikeouts \times 2)/IP + FIP\ constant. \quad (1)$$

As discussed in the introduction, we would like to calculate opponent FIP for each game a team plays. This statistic is not readily available from a box score or game log anywhere. We will first describe how we obtain each quantity in formula (1).

The number of home runs, walks/hit by pitch, and strikeouts can be found from Baseball Reference's team batting game logs.[10] The FIP constant each year is recorded by Fangraphs.

The number of innings that the opponent pitched is not recorded in a batting game log. It is not always 9 innings. If the home team is winning after 8½ innings, they will not bat in the bottom of the 9th and the away team will pitch 8 innings. If the game ends in a walk-off, the away team will have finished the game having pitched only part of the last inning. If the game goes into extra innings, both teams will pitch more than 9 innings. Here's how we find the number of innings pitched:

1. Read in the pitching game logs from every game from that season. We call the aforementioned team Team X, and their opponent Team Y.
2. Match Team X's batting game log with Team Y's pitching game log from the same game. A pitching game log contains how many innings each of their pitchers pitched.
3. Add up the number of innings each of Team Y's pitchers pitched, to find the total number of innings pitched.

To summarize, we record:

- The number of home runs their batters hit
- The number of walks/hit by pitch for their batters
- The number of times their batters have struck out
- The number of innings the opponent pitchers collectively pitched
- The FIP constant for that year

We repeat this procedure for every game a team plays and for all 30 teams. Using this, we calculate opponent FIP for each game for each team.

# 3 OBSERVATION OF A UNIVERSAL PATTERN

Each team has a data set of approximately 162 opponent FIP values(each team is scheduled to play 162 games in a season). We want to examine if there was anything in common between each team's data sets. We will use data from the 2019 MLB season as an illustration.

One possible approach is to plot the histogram of the opponent FIP values for each team according to the frequency of each value. However, a histogram using the exact values would not be practical -- most FIP values would only have one or two data points, as it is uncommon to get a single opponent FIP value more than a couple of times in a season, given that they can realistically range anywhere from -1 to around 15. As such, we would have to divide FIP values into a set of bins, with all points falling inside a bin grouped together, and plot the histogram for the bins. Doing so is sensible as FIP values of 3 and 3.1 reflect more or less a very similar level of performance.

The choice of size for the bin is delicate: it should not be too small such that only a small number of data points fall into it and it should not be too large -- clearly FIP values of 3 and 4 indicate different levels of performance. After some experiments, we find that 0.5 appears to be one of the best possible bin sizes. The histogram for the Yankees' data is shown below.

*Graph 1: Histogram for the 2019 Yankees Opponent FIP Values*

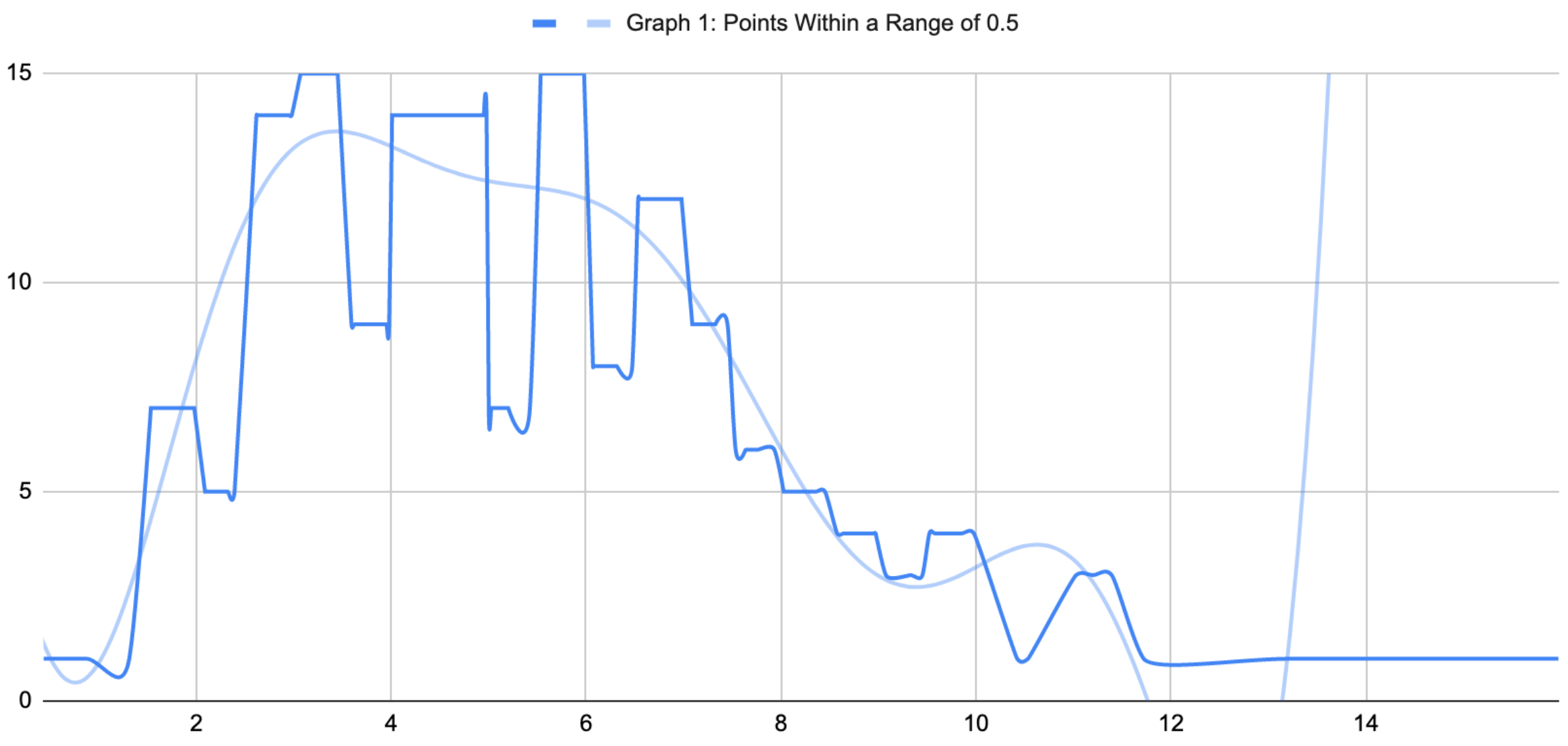

Even though this bin value worked better than others, the data still does not have a consistent shape and fluctuates between higher and lower points. It only very loosely follows the curve of best fit. This indicates that there is not a large enough sample size, and we cannot compare teams using their distribution graphs.

Although these histograms may provide valuable information, for our purposes it is more suitable to use alternative methods.

We instead studied correlations between data sets directly. There are various techniques for this, such as the Kolmogorov-Smirnov test, and qq-plots along with Pearson correlation. The simplest is qq-plots, which we will use here as an illustration. However, the discussion can readily be extended to other methods.

To create the plot, the two selected data sets are ordered. Each point on the plot has an x and y coordinate. Its x coordinate corresponds to a data point in one set, and its y coordinate corresponds to a data point in another set with the same percentile. The more linear the qq-plot is, the more similar the distribution of the two sets is.

Each team has a data set with ~162 opponent FIP points. In order to compare the distributions of all of the teams, we created a data set for a hypothetical league average team. This was done in the following way:

1. Combine all of the data sets of each team into one large data set with ~4860(162 games × 30 teams) terms.

2. This set of 4860 numbers can be ordered from least to greatest and sorted into 162 groups. If $n_m$ is the mth smallest term in the set, then the groups are $n_1$ to $n_{30}$, $n_{31}$ to $n_{60}$ … $n_{4831}$ to $n_{4860}$ such that each group contains 30 numbers.
3. The *middle value* (15th smallest term) in each group is then selected to be put in a new set of 162 terms. This new set can be used to form qq-plots with the teams who played 162 games.
4. Note: Some seasons(such as 2019) will have slightly more or less than 4860 games, because a couple of teams may play 161 or even 163 or 164 games. However, most seasons just have 4860, and any season with less than 4855 games or more than 4865 games is extremely unusual. If a season does not have 4860 total games, the procedure is essentially the same. Making the adjustment is quite trivial.

If the qq-plot is the line x=y, then the two distributions are identical. If it is linear but does not pass through the origin or has a slope not equal to 1, then the distribution of one data set is just a stretched, shifted version of the other. But the shape of the distribution is the same. If the qq-plot is not linear, the distributions of the data sets are very different. A few examples of qq-plots between teams' data sets and the hypothetical league average team's data set are shown below.

*Graph 2: QQ-Plot of the 2019 Red Sox Against the Hypothetical League Average Team*

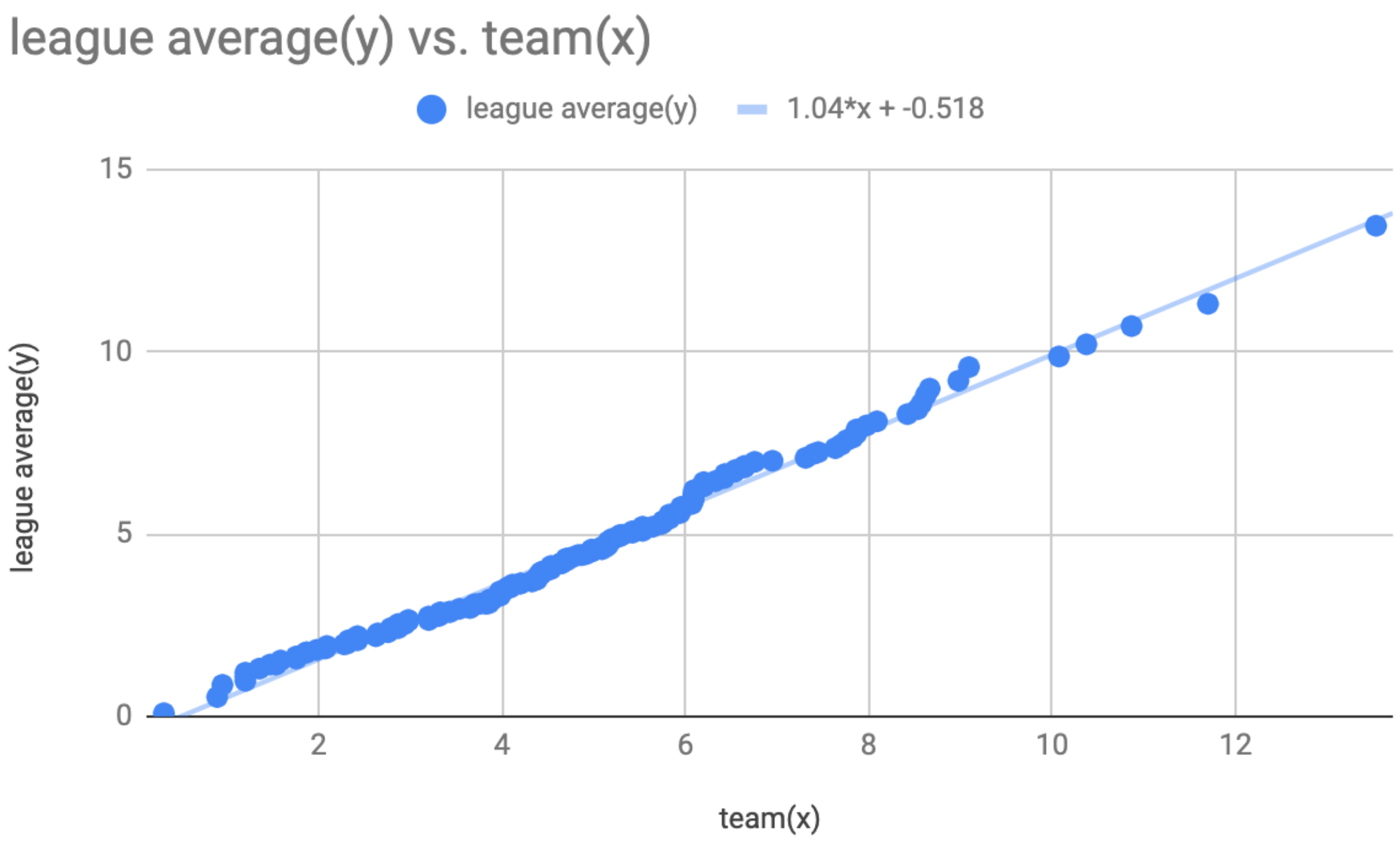

*Graph 3: QQ-Plot of the 2019 Angels Against the Hypothetical League Average Team*

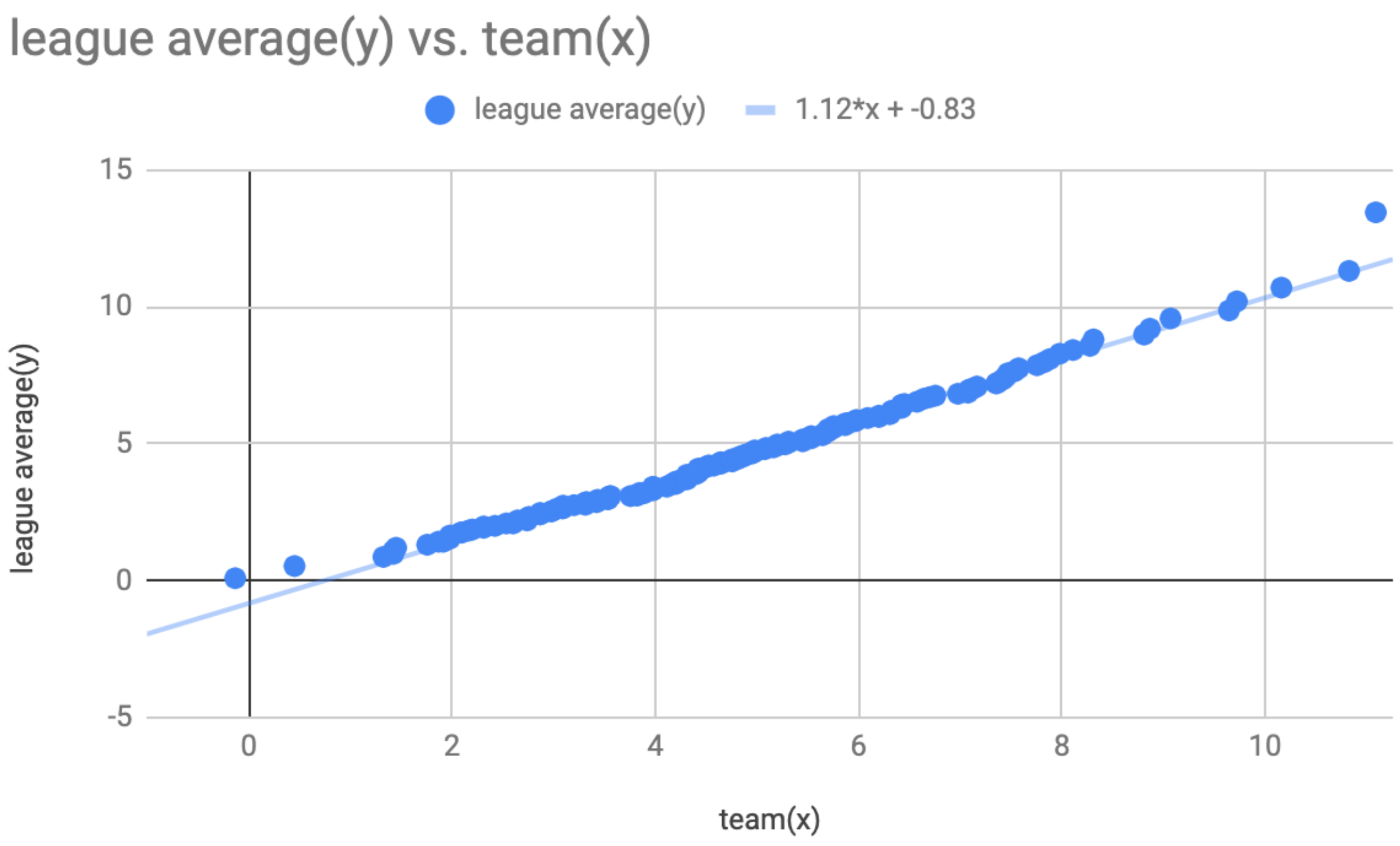

*Graph 4: QQ-Plot of the 2019 Cubs Against the Hypothetical League Average Team*

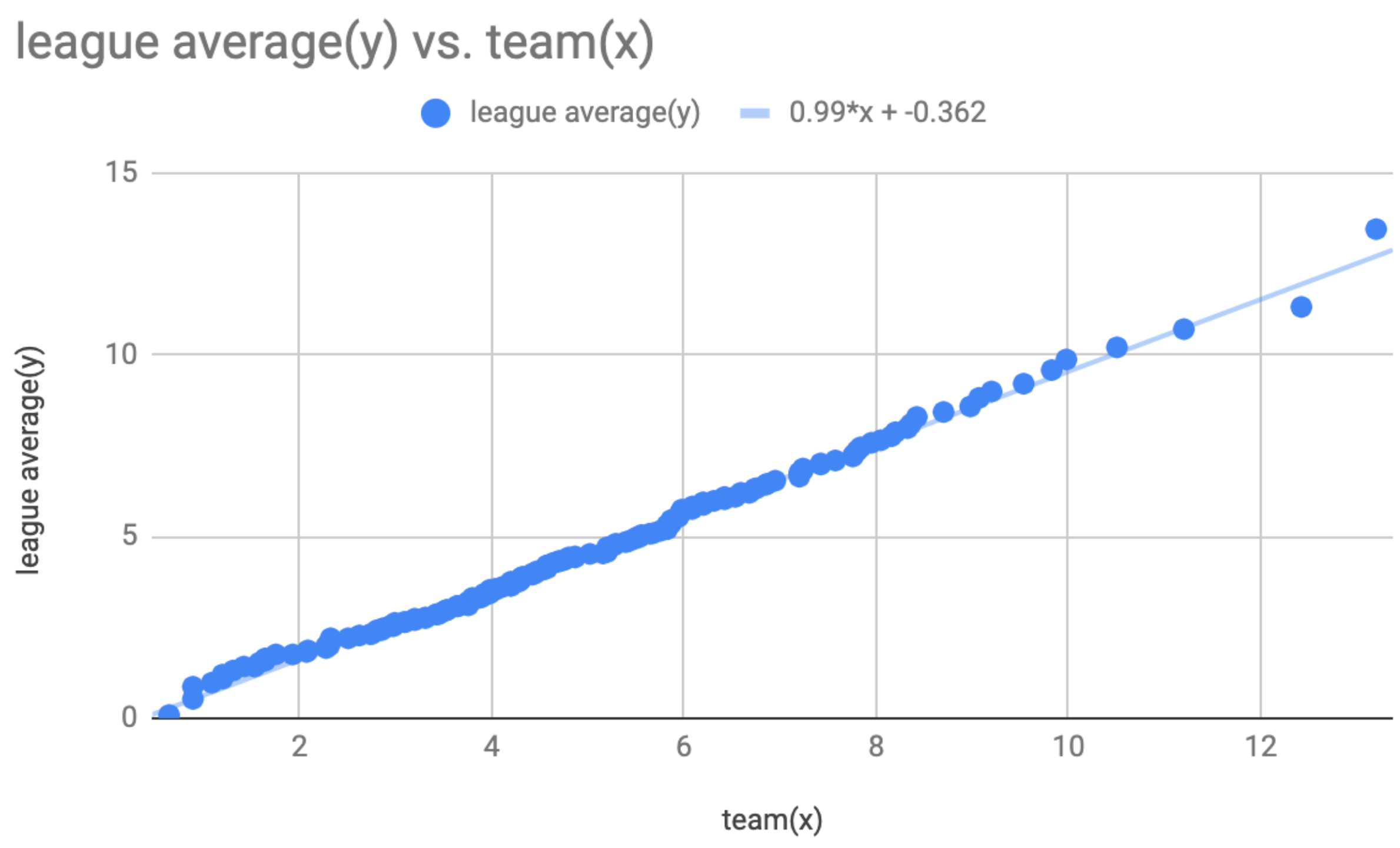

As one can see, the qq-plots are very close to lines. In order to quantify the extent to which the qq-plots are linear, we calculated the Pearson Correlation Constant for each of the qq-plots. The results are shown in the table below.

The *Pearson Correlation Constant* is used to measure the strength of correlation of a data set. In other words, it measures how closely the data points follow the line of best fit. This is useful here to quantify just how similar the teams' data distributions are. The higher the constant is, the more linear the qq-plot is, and the more similar the distributions are.

*Table 2: Pearson Constants of QQ-Plots*

| Team | Pearson |
|---|---|
| ARI | 0.9887 |
| ATL | 0.9889 |
| BAL | 0.9884 |
| BOS | 0.9889 |
| CHC | 0.9914 |
| CHW | 0.9915 |
| CIN | 0.9857 |
| CLE | 0.9889 |
| COL | 0.9895 |
| DET | 0.9836 |
| HOU | 0.9915 |
| KCR | 0.9903 |
| LAA | 0.9885 |
| LAD | 0.9877 |
| MIA | 0.9914 |
| MIL | 0.9878 |
| MIN | 0.9924 |
| NYM | 0.9895 |
| NYY | 0.9927 |
| OAK | 0.9908 |
| PHI | 0.9818 |
| PIT | 0.9893 |
| SDP | 0.9894 |
| SEA | 0.9896 |
| SFG | 0.9908 |
| STL | 0.9878 |
| TBR | 0.9900 |
| TEX | 0.9920 |
| TOR | 0.9844 |
| WSN | 0.9843 |
| Average | 0.9889 |

Pearson values can be anywhere from -1 to 1. The closer their absolute value is to 1, the higher the correlation is.

The high Pearson values tell us that the qq-plots are very close to straight lines, which is significant because it means that every team's opponent FIP data set's distribution is just a shifted, stretched version of the hypothetical league average team's distribution. The different ways each team's distribution is stretched or shifted indicates certain qualities of the team's offensive ability.

A team whose data is very condensed is consistent in their offensive output from game to game. A team whose data is shifted towards higher numbers is one with a high average offensive output. For example, the 2019 Houston Astros had an average opponent FIP of 5.78 with a standard deviation of 2.52. The 2019 Los Angeles Dodgers had an average of 5.47 and a standard deviation of 2.87. The Astros were a better offensive team on average, and their performance was more reliable. The Dodgers were worse offensively, but their offensive output was more unpredictable. They were more likely to put up an exceptional offensive game than the Astros.

# 4 APPLICATIONS

This finding, that each team's opponent FIP distribution is extremely similar in shape, should have many applications. Here, we discuss a particularly important one for MLB pitchers.

As mentioned in the introduction, one of the major problems for pitchers in baseball is that no matter which team they face, their ERA and FIP are counted the same way. Pitching seven innings and giving up three runs against a top offensive team is a more impressive task than doing the same against a weak team, but they are counted as the same.

We will now describe an algorithm to take advantage of this universal distribution and create a new player evaluation metric. Since each team's distribution has roughly the same shape up to stretching and shift, we propose to equate points from each data set that have the same percentile. For example, one could equate the 15th percentile point on the Yankees' data to the 15th percentile point on the Diamondbacks' data. This would help solve the strength of schedule problem, as then one could compare a start against the Yankees with a start against the Diamondbacks. Therefore, we propose a statistic called aFIP. aFIP measures what a pitcher's FIP would have been if he had faced a league average team in every game he pitched. aFIP is calculated game by game in the following way:

1. Let's say that in the selected game, the pitcher faced Team E. One then finds the percentile of the selected pitcher's FIP among Team E's data set (the data points in each team's opponent FIP data are combined FIP values of all of the pitchers during a game, but the selected pitcher's FIP value is just his own value, and not a combination).
2. Then, one finds the value among the hypothetical league average team's data that has the same percentile. That value is the pitcher's aFIP. This is the equipercentile method of calculating aFIP. For example, suppose a player had an FIP of 3.54 against Team E, and that was in their 37th percentile. His aFIP would be the 37th percentile value on the League Average distribution. That value is 3.58, so his aFIP using this method is 3.58.

Now, observe that the Pearson values were not exactly 1, meaning that the data points on the qq-plot did not exactly follow the line of best fit. This leads to some possible uncertainties in the calculation of aFIP.

If the Pearson constant was equal to 1 on the qq-plots of all 30 MLB teams, then we could simply use the equation of the line that was created by the data points to calculate aFIP. For example, if the equation of the line was y=1.17x-0.35 (where x is the FIP value from the MLB team's data and y is the FIP value from the hypothetical league average term), then if a pitcher got a 2.40 FIP against that team, his aFIP would be y=1.17(2.40)-0.35 → y=2.458 → 2.46.

We call this method of calculating aFIP the Slope-Intercept Method.

Here are the aFIP results from the Slope-Intercept Method compared to the results of the equipercentile aFIP method using season stats from the 61 qualified pitchers in 2019:

*Table 3: aFIP and aFIP Difference of Qualified Starting Pitchers in 2019*

| player | slopeIntercept | equipercentile | FIP | SI difference | EQP difference | EQP - SI |
|---|---|---|---|---|---|---|
| Gerrit Cole | 2.65 | 2.73 | 2.64 | 0.01 | 0.09 | 0.08 |
| Jacob deGrom | 2.67 | 2.64 | 2.67 | 0.00 | -0.03 | -0.03 |
| Lance Lynn | 2.92 | 2.88 | 3.13 | -0.21 | -0.25 | -0.04 |
| Max Scherzer | 2.49 | 2.40 | 2.45 | 0.04 | -0.05 | -0.09 |
| Justin Verlander | 3.26 | 3.28 | 3.27 | -0.01 | 0.01 | 0.02 |
| Charlie Morton | 2.69 | 2.67 | 2.81 | -0.12 | -0.14 | -0.02 |
| Stephen Strasburg | 3.28 | 3.30 | 3.25 | 0.03 | 0.05 | 0.02 |
| Shane Bieber | 3.35 | 3.41 | 3.32 | 0.03 | 0.09 | 0.06 |
| Zack Greinke | 3.15 | 3.18 | 3.22 | -0.07 | -0.04 | 0.03 |
| Lucas Giolito | 3.27 | 3.30 | 3.43 | -0.16 | -0.13 | 0.03 |
| Walker Buehler | 3.04 | 3.03 | 3.01 | 0.03 | 0.02 | -0.02 |
| Hyun-Jin Ryu | 3.06 | 3.11 | 3.1 | -0.04 | 0.01 | 0.05 |
| Patrick Corbin | 3.55 | 3.59 | 3.49 | 0.06 | 0.10 | 0.04 |
| Jack Flaherty | 3.39 | 3.39 | 3.46 | -0.07 | -0.07 | 0.00 |
| Zack Wheeler | 3.55 | 3.51 | 3.48 | 0.07 | 0.03 | -0.04 |
| Jose Berrios | 4.12 | 4.18 | 3.85 | 0.27 | 0.33 | 0.06 |
| Noah Syndergaard | 3.67 | 3.71 | 3.6 | 0.07 | 0.11 | 0.04 |
| Sonny Gray | 3.38 | 3.32 | 3.42 | -0.04 | -0.10 | -0.06 |
| Mike Minor | 4.02 | 4.01 | 4.25 | -0.23 | -0.24 | -0.01 |
| Kyle Hendricks | 3.57 | 3.52 | 3.61 | -0.04 | -0.09 | -0.05 |
| Luis Castillo | 3.63 | 3.69 | 3.7 | -0.07 | -0.01 | 0.05 |
| Mike Soroka | 3.52 | 3.47 | 3.45 | 0.07 | 0.02 | -0.05 |
| Marcus Stroman | 3.75 | 3.81 | 3.72 | 0.03 | 0.09 | 0.06 |
| Eduardo Rodriguez | 3.90 | 4.00 | 3.86 | 0.04 | 0.14 | 0.10 |
| Marco Gonzales | 3.99 | 3.97 | 4.15 | -0.16 | -0.18 | -0.02 |
| Jose Quintana | 3.76 | 3.89 | 3.8 | -0.04 | 0.09 | 0.12 |
| Aaron Nola | 4.02 | 3.99 | 4.03 | -0.01 | -0.04 | -0.03 |
| German Marquez | 4.08 | 4.09 | 4.06 | 0.02 | 0.03 | 0.00 |
| Clayton Kershaw | 3.89 | 3.93 | 3.86 | 0.03 | 0.07 | 0.04 |
| Joe Musgrove | 3.74 | 3.67 | 3.82 | -0.08 | -0.15 | -0.08 |
| Matthew Boyd | 4.31 | 4.39 | 4.32 | -0.01 | 0.07 | 0.08 |
| Trevor Bauer | 4.49 | 4.47 | 4.34 | 0.15 | 0.13 | -0.03 |
| Masahiro Tanaka | 4.31 | 4.23 | 4.27 | 0.04 | -0.04 | -0.08 |
| Madison Bumgarner | 3.89 | 3.86 | 3.9 | -0.01 | -0.04 | -0.03 |
| Max Fried | 3.75 | 3.89 | 3.72 | 0.03 | 0.17 | 0.14 |
| Homer Bailey | 4.17 | 4.12 | 4.11 | 0.06 | 0.01 | -0.05 |
| Jon Lester | 4.25 | 4.20 | 4.26 | -0.01 | -0.06 | -0.05 |
| Yu Darvish | 4.38 | 4.57 | 4.18 | 0.20 | 0.39 | 0.19 |
| Miles Mikolas | 4.32 | 4.24 | 4.27 | 0.05 | -0.03 | -0.08 |
| Anibal Sanchez | 4.60 | 4.63 | 4.44 | 0.16 | 0.19 | 0.03 |
| Anthony DeSclafani | 4.41 | 4.51 | 4.43 | -0.02 | 0.08 | 0.09 |
| Robbie Ray | 4.33 | 4.23 | 4.29 | 0.04 | -0.06 | -0.10 |
| Sandy Alcantara | 4.41 | 4.50 | 4.55 | -0.14 | -0.05 | 0.10 |
| Reynaldo Lopez | 5.28 | 5.49 | 5.04 | 0.24 | 0.45 | 0.21 |
| Brad Keller | 4.61 | 4.69 | 4.35 | 0.26 | 0.34 | 0.08 |
| Adam Wainwright | 4.32 | 4.40 | 4.36 | -0.04 | 0.04 | 0.08 |
| Joey Lucchesi | 4.23 | 4.16 | 4.17 | 0.06 | -0.01 | -0.06 |
| Merrill Kelly | 4.64 | 4.63 | 4.51 | 0.13 | 0.12 | -0.02 |
| Ivan Nova | 5.08 | 5.09 | 4.98 | 0.10 | 0.11 | 0.02 |
| Brett Anderson | 4.49 | 4.48 | 4.57 | -0.08 | -0.09 | -0.01 |
| Wade Miley | 4.53 | 4.46 | 4.51 | 0.02 | -0.05 | -0.07 |
| Tanner Roark | 4.62 | 4.57 | 4.67 | -0.05 | -0.10 | -0.06 |
| Martin Perez | 4.92 | 4.93 | 4.66 | 0.26 | 0.27 | 0.01 |
| Rick Porcello | 4.81 | 4.79 | 4.76 | 0.05 | 0.03 | -0.02 |
| Mike Fiers | 4.85 | 4.67 | 4.97 | -0.12 | -0.30 | -0.17 |
| Julio Teheran | 4.92 | 5.01 | 4.66 | 0.26 | 0.35 | 0.09 |
| Jakob Junis | 4.88 | 4.91 | 4.82 | 0.06 | 0.09 | 0.03 |
| Zach Eflin | 5.00 | 5.15 | 4.85 | 0.15 | 0.30 | 0.15 |
| Jeff Samardzija | 4.58 | 4.62 | 4.59 | -0.01 | 0.03 | 0.04 |
| Dakota Hudson | 4.94 | 5.00 | 4.93 | 0.01 | 0.07 | 0.06 |
| Mike Leake | 5.26 | 5.39 | 5.19 | 0.07 | 0.20 | 0.12 |

*Note: FIP is conventionally displayed to the hundredths place*

As can be seen in the table, many pitchers have significant aFIP-FIP differences. Eight pitchers have a difference above 0.20 for both methods of calculating aFIP, which is a very significant amount. Furthermore, the average absolute difference between the aFIP values from the SI and EQP methods is 0.060, meaning that the two methods are relatively closely aligned.

# 5 IMPORTANCE OF AFIP

aFIP can impact player evaluation greatly, as for some pitchers the difference between aFIP and FIP was as large as 0.31. This is a very significant difference, as the pitcher who finished second in the NL Cy Young voting in 2019(Hyun-Jin Ryu) had an FIP that was only 0.39 greater than the pitcher who finished 11th(Patrick Corbin).

Such a change can make a dramatic difference in award voting, all-star nominations, and teams' decisions at the executive level. A team looking for a starting pitcher may be considering Jose Berrios(3.85 FIP) and Joe Musgrove(3.82 FIP) as two of their trade options, without a clear favorite. However, knowing that Berrios's aFIP is 4.15 while Musgrove's aFIP is 3.69 may provide some more insight. More accurate player evaluation results in greater team success.

Baseball is a game of inches, where even the slightest improvement could make a huge difference. The last four world series champions(as of the end of the 2019 season) have played a total of 11 elimination games during the playoffs, and as Cubs executive Theo Epstein put it, "If Zobrist's ball is three inches farther off the line, I'm on the hot seat for a failed five-year plan." Picking the right pitcher could make all the difference in a World Series run. Winning a World Series or even just a couple extra playoff games may have a significant impact on a team financially, as a World Series run could earn a team anywhere from $50 million to $250 million 15 years ago, while a single playoff game could earn a team around $6.8 million (Matheson and Blade, 2005).

The amount of money paid to a pitcher could be changed by such a statistic as well. If a team knows that a player's FIP was inflated by 0.2 due to the fact that he faced weaker opponents, they may be less inclined to pay him as much money in free agency. That 0.2 could make a huge difference. For example, when Yu Darvish hit free agency in the 2017-18 offseason, he had a 3.57 FIP over his past two seasons. Even with his previous injury history which caused him to only appear in an average of 17.5 games over the last four seasons, he was signed to a six-year, $126 million contract. On the other hand, Jake Arrieta hit free agency in the same year after a 3.81 FIP over his last two seasons. He had previously won a Cy Young award and had no such injury history, appearing in an average of 29.75 games over his last four seasons. Both pitchers are about the same age(Arrieta is about half a year older). Arrieta signed a three-year, $75 million deal.

While Arrieta got a higher average annual value for this initial deal, Darvish's contract is considered much better. Even at the time, it was widely expected that once Arrieta hit free agency at the end of his age-35 season, it would be unexpected for him to make more than $51 million after his three-year, $75 million deal ended(51+75 = 126, Darvish's number). That ended up being true, as Arrieta signed just a 1-year, $6 million deal following the conclusion of that contract. Almost all MLB players prefer long-term contracts because all of the money is fully guaranteed -- and due to the financial security -- and that 0.24 difference in FIP was likely one of the factors examined when Darvish and Arrieta got their contracts.

Due to its ability to take into account pitchers' difference in schedule strength -- which we have shown is a significant influence on pitchers' FIP -- aFIP is a statistic that would be useful in player evaluation. aFIP shift is also consistent with strength of schedule. For example, as we noted earlier,

Jose Berrios faced a very weak schedule while Charlie Morton faced a very difficult schedule. Berrios' FIP was 3.85, while his aFIP values were 4.12(SI) and 4.18(EQP). Morton's FIP was 2.81, and his aFIP values were 2.69(SI) and 2.67(EQP).

# 6 DISCUSSION

aFIP provides a way to evaluate pitchers based on an already established statistic in FIP. It allows two pitchers who played schedules with different degrees of strength to be fairly compared.

aFIP does not have complete accuracy. The Pearson constants of the qq-plots, while very close to 1, were not exactly 1. This resulted in some small differences between the values calculated by the Slope-Intercept and Equipercentile methods.

The observation of the universality and similar types of equipercentile algorithms apply to other seasons as well, along with other stats like ERA. Furthermore, using the distributions, we can also attempt to generate statistically rigorous evidence for artificial tampering or cheating.

There are a couple of subtle points that need to be addressed.

One obvious concern is the following: If a player's stats can be skewed by an unfair schedule, couldn't a team's statistics be skewed due to the same logic? This is a valid concern. However, at first glance, it is clear that it is much less of a concern than pitchers' schedules because each team's schedule follows a specific structure from the MLB. Each team plays 19 games against each of their division opponents(76 total) and 6 or 7 games against all other teams in their league(66 total) that are not in their division. Obviously, if one division had far stronger pitching than another, it could cause the results to be skewed. But there is far less randomness in a team's schedule than in a pitcher's schedule.

The average opponent for an AL East team gives up 4.83 runs per game, 4.87 for an AL Central team, and 4.86 for an AL West team, excluding interleague games. The average opponent for an NL East team gives up 4.76 runs, 4.73 for an NL Central team, and 4.75 for an NL West team. A team only plays 20 interleague games per year against 6 random opponents, so those games should have very little impact, especially on pitchers, who only pitch around 3-5 interleague games per year.

In sum, differences in strength of schedule for teams are very small, and these minimal differences make no noticeable difference in the calculation of aFIP.

Another subtle point that needs addressing is the following:

Although the differences between the aFIP values calculated by the equipercentile and slope intercept methods are small -- 40 of the 61 pitchers have a difference of 0.06 or less -- it is possible that a small difference could come from points very far from the line “evening out”. For example, if the SI-EQP difference was 1.5 for one game and -1.5 for another, it would appear that the two methods are closely aligned.

That being said, none of the 61 pitchers had a very extreme SI-EQP difference. No pitcher has a difference greater than 0.21 and only three have a difference greater than 0.15.

Perhaps more advanced metrics could address this issue more thoroughly, but it is likely not a great concern.

Calculating aFIP for pitchers’ 2019 seasons is just one very specific application of the discovery that teams’ distributions are very similar to each other. The principle holds true for other MLB seasons as well, and for other statistics such as ERA.

We can use aFIP distributions to attempt to make a statistically backed claim of artificial tampering. For example, the Houston Astros were under investigation for illegally stealing signs to boost their offensive performance during home games. Plotting home games against away games on a qq-plot yielded a very linear graph, with a Pearson constant of 0.9906. This eliminates one possible indicator of the Astros cheating. Although they eventually admitted to having cheated, this would have still been useful information in the investigation. Also, although the Astros’ cheating did not show up with a discrepancy in the distribution of their opponent FIP data, it is possible that future cases of artificial tampering will.

---

Like just about every pitching measurement in existence, aFIP should not necessarily be treated as an exact science. But it is a useful and accurate statistic that accounts for a factor no existing stats have yet, provides a solid reference point, and is a valuable tool for player evaluation.

This finding of a universal pattern has even more applications and extensions that can be further developed in the future.